\documentclass[pra,twocolumn,nofootinbib,preprint,amssymb,longbibliography,amsfonts,amsmath,superscriptaddress,showpacs]{revtex4-1}
\usepackage{adjustbox,lipsum}
\usepackage{graphicx}
\usepackage{bm}
\usepackage{color}
\usepackage{url}
\usepackage{epstopdf}
\usepackage{amsthm,mathrsfs} 
\usepackage{txfonts}
\usepackage[colorlinks=true,linkcolor=blue,urlcolor=blue,citecolor=blue,pdfusetitle]{hyperref}

\usepackage{newtxtext}
\usepackage[varvw]{newtxmath}

\usepackage[framemethod=TikZ]{mdframed}
    
    \usepackage{tikz}
\usetikzlibrary{calc}

\xdefinecolor{mycolor}{RGB}{62,96,111} 
\colorlet{bancolor}{mycolor}

\mdfdefinestyle{MyFrame}{%
    linecolor=black,
    outerlinewidth=2pt,
    innertopmargin=4pt,
    innerbottommargin=4pt,
    innerrightmargin=4pt,
    innerleftmargin=4pt,
        leftmargin = 4pt,
        rightmargin = 4pt,
    backgroundcolor=gray!50!white
        }

\begin{document}
\count\footins = 1000
\title{
{Non-linear media in weakly curved spacetime: optical solitons and probe pulses for gravimetry}}

\author{Alessio Belenchia}
\affiliation{Institut f\"{u}r Theoretische Physik, Eberhard-Karls-Universit\"{a}t T\"{u}bingen, 72076 T\"{u}bingen, Germany}
\affiliation{Centre for Theoretical Atomic, Molecular, and Optical Physics, School of Mathematics and Physics, Queens University, Belfast BT7 1NN, United Kingdom}
\author{Felix Spengler}
\affiliation{Institut f\"{u}r Theoretische Physik, Eberhard-Karls-Universit\"{a}t T\"{u}bingen, 72076 T\"{u}bingen, Germany}
\author{Dennis R\"{a}tzel}
 \affiliation{ZARM, University of Bremen, Am Fallturm 2, 28359 Bremen, Germany}
\affiliation{Humboldt  Universit\"{a}t  zu  Berlin,  Institut  f\"{u}r  Physik, Newtonstraße  15,  12489  Berlin,  Germany}
\author{Daniel Braun}
\affiliation{Institut f\"{u}r Theoretische Physik, Eberhard-Karls-Universit\"{a}t T\"{u}bingen, 72076 T\"{u}bingen, Germany}

\date{\today}

\begin{abstract}
That light propagating in a gravitational {field} 
gets frequency-shifted is one of the basic consequences of any metric theory of gravity rooted in the equivalence principle. At the same time, also a time dependent material's refractive index can frequency-shift light propagating in it. The mathematical analogy between the two effects is such that the latter has been used to study the optical analogue of a black-hole spacetime. 
{Here, we combine these two effects by showing that light propagation in non-linear media in the presence of a moving refractive index perturbation can lead to a gravity-dependent blueshift. We find that the predicted blueshift surpasses the gravitational redshift even if the medium is considered to be perfectly stiff. In realistic scenarios, by far the strongest frequency shift arises due to the deformation of the dielectric medium and the corresponding photoelastic change of refractive index. This has the potential to facilitate optical sensing of small gravity gradients.
}
\end{abstract}
\maketitle

\section{Introduction}
Electromagnetism in media and electromagnetism in curved spacetime, while seemingly disparate topics, have been shown to be surprisingly connected. 
The pioneering works of Gordon and Plebanski~\cite{gordon1923lichtfortpflanzung,plebanski1960electromagnetic} have shown the equivalence of light propagation in curved spacetime and in optical media. These results have been used to investigate the analogue of exotic gravitational effects, such as the Hawking radiation and cosmological particle creation, in optical laboratory systems~\cite{philbin2008fiber,westerberg2014experimental,rubino2011experimental,PhysRevLett.105.203901,PhysRevLett.122.010404}. At the same time, at the level of applied physics, the methods and tools of general relativity have shaped the field of transformation optics~\cite{leonhardt2011invisibility,chen2010transformation,leonhardt2009transformation}. 

Of particular interest {to our endeavour} are a series of works showing how optical solitons in non-linear dielectric media give rise to a refractive index perturbation (RIP) which, from the point of view of weaker co-propagating probe pulses, is equivalent to a white-hole horizon. First proposed by Philbin et al.~\cite{philbin2008fiber}, these systems have been the focus of intense investigation for a variety of purposes, from analogue gravity to all-optical amplification schemes relying on relativistic scattering~\cite{westerberg2014experimental,rubino2011experimental,PhysRevLett.105.203901,PhysRevLett.122.010404,rubino2012negative,faccio2012optical,rubino2012soliton,petev2013blackbody,roger2013high,belgiorno2010quantum}. In a nutshell, weak probe pulses co-propagating with a soliton experience an increase in the local refractive index while approaching it. By choosing parameters in an appropriate way, and accounting for the dispersive features of the medium, it is possible to obtain a scattered weak pulse that has been strongly blueshifted in frequency and that lags behind the propagating soliton.

Scattering of light from a relativistic RIP has been studied in detail considering light propagating in non-linear media in flat spacetime~\cite{rubino2012soliton}. {Building upon the equivalence between light propagation in curved spacetime and in a medium, in~\cite{spengler2023optical} we have shown how a weak gravitational field affects the propagation of an optical soliton.} 
In this work, we {consider}
the effect that a weak gravitational field has on the blueshift induced on a weak probe pulse by the RIP due to the soliton. {We do this by comparing the frequency shift experienced by probe pulses propagating along different trajectories and show that }{such an effect} {is larger than gravitational redshift already when considering a perfectly stiff material. For deformable materials, the action of Newtonian gravity induces additional refractive index perturbations, a.k.a. photoelasticity, which can render the effect orders of magnitude larger than the gravitational redshift. 
This last observation has the potential to aid in probing weak gravitational fields via optical experiments.}

{The work is structured as follows. In Sec.II, we review the analogy between light propagating in an optical medium at rest in a static spacetime and the propagation in an effective optical medium in flat spacetime~\cite{spengler2023optical}.}
We then consider the RIP as a linear propagating refractive index inhomogeneity in the effective optical medium and solve for the trajectory of the probe pulse in Sec.III. Comparing propagation at constant radius and radially propagating pulses, we show that the blueshift experienced by weak probe pulses depends on the gravitational acceleration
and that the magnitude of such an effect exceeds the one of standard gravitational redshift. 
In Sec.IV, we account for this large blueshift effect
by approximate momentum conservation considerations and, finally, we conclude in Sec.V with discussion and outlooks.

\section{Effective medium analogy in geometric optics}
The formal analogy between Maxwell's equations in curved spacetime and Maxwell's equations in an optical medium in flat spacetime has been known since the seminal work of Plebanski~\cite{plebanski1960electromagnetic}. Here, we consider an aspect of this analogy, explicitly stated in our recent work~\cite{spengler2023optical}, that connects Maxwell's equations in an optical medium stationary in curved spacetime (specifically Schwarzschild spacetime) to Maxwell's equations in an effective medium in flat spacetime.
We review this analogy here in the geometric optic limit since we will be interested in describing light rays propagating, in the presence of an RIP, in non-linear materials stationary in Schwarzschild spacetime. {From now on, unless specified differently, we assume $c=1$.}

\subsection{Hamiltonian formalism}

Light propagation in curved spacetime and in the presence of a medium in the geometric optic limit has been the subject of many works; see e.g.~\cite{perlick2000ray} and references therein. Here we follow the discussion in~\cite{1975A&A....44..389B,perlick2000ray}. 
Consider a medium whose rest frame is identified by the observer vector field $U^{\mu}$. By introducing the light wavevector $p_\mu=\nabla_\mu\phi$, defined in terms of the gradient of the phase of the light wave\footnote{The symbol $\nabla_{\mu}$ indicates the covariant derivative.}, we can define the frequency as measured in the rest frame of the medium
\begin{equation}
    \omega=-p_\mu U^\mu,
\end{equation}
where we use signature $-\!+\!++$ for the spacetime metric $g_{\mu\nu}$.

The phase velocity of the light seen by a generic observer $O^{\mu}$ is given by the invariant relation
$1/\bar{v}_{\rm ph}^2=1+{p\cdot p}/(p\cdot O)^2$. The refractive index is defined as the inverse of this phase velocity in the rest frame of the medium, i.e., {choosing $O^\mu=U^{\mu}$}, leading to $n^2=1+{p\cdot p}/{(p\cdot U)^2}$. 
This dispersion relation can be rewritten in the form $\mathcal{H}=0$ by defining 
\begin{equation}\label{H0}
    \mathcal{H}=\frac{1}{2}(g^{\mu\nu}-(n^2-1)U^\mu U^\nu)p_\mu p_\nu.
\end{equation}
The characteristic curves of this Hamiltonian are the light rays and describe the propagation of light in the geometric optics limit\footnote{As it is well known, in the absence of dispersion, i.e. {when the refractive index $n$ is independent of the frequency}, the Gordon metric can be readily obtained by rewriting $\mathcal{H}=\frac{1}{2}(g^{\mu\nu}-(n^2-1)U^\mu U^\nu)p_\mu p_\nu=\frac{1}{2}(\tilde{g}_{\rm Gordon}^{\mu\nu}(x))p_\mu p_\nu$. In this case, the light rays are null geodesics of the Gordon optical metric. When dispersion is present, the Gordon metric is not a Lorentzian metric and the Hamiltonian equations of $\mathcal{H}$ do not correspond to the geodesic flow of an optical metric. Nevertheless, the full power of the Hamiltonian framework can still be used to find the rays and the evolution of the frequency~\cite{1975A&A....44..389B}.}.
In~\cite{1975A&A....44..389B}, these equations are investigated in detail and the corresponding equation for the light's frequency is derived. We report here this \textit{redshift equation} for completeness\footnote{{It should be noted that this approach, when dealing with dispersive media, improves the one sported in~\cite{cacciatori2010spacetime}. For a brief discussion see Appendix~\ref{appendix0}.}}

{\begin{widetext}
\begin{equation}\label{redshiftEq}
    \frac{d (-p_a U^a)}{d{s}
    }=-(p_a U^a)^2\left[\left(\sigma_{\alpha\beta}N^\alpha N^\beta+\frac{1}{3}\theta\right) n+a_\mu{N}^\mu+\nabla_{\mu} n {U}^\mu\right] n, 
\end{equation}
\end{widetext}
where $s$ is an affine parameter, $a_\mu=U^\nu\nabla_\nu U_\mu$ is the four-acceleration, $\sigma_{\alpha\beta}$ the symmetric shear, and $\theta$ the expansion of the vector field $U^\mu$, while $N^\mu$ is a unit vector field orthogonal to $U^\mu$ and $h_{\mu\nu}=g_{\mu\nu}+U_\mu U_\nu$ is the projector on the three-space orthogonal to $U^\mu$.} 

\subsection{Effective medium in flat spacetime}

Let us now consider a spherically symmetric spacetime in isotropic coordinates. The metric can be written in full generality as 
\begin{equation}\label{metricspehrical}
    d{\rm s}^2=-\left(\frac{B}{A}\right)^2 dt^2+A^4 a^2\delta_{\alpha\beta} dx^\alpha dx^\beta,
\end{equation}
with $A=A(r,t),\,B=B(r,t),\,a=a(t)$, and $r=\sqrt{x^2+y^2+z^2}$. 
Considering the observer vector field $U^\mu=\delta^{\mu 0}/||U||$, a simple manipulation of 
Eq.~\eqref{H0} shows that 
\begin{align}
    \mathcal{H}=\frac{1}{2}\Omega \left(\left[\eta^{\mu\nu}-(\tilde{n}^2-1)\right]V^{\mu} V^{\nu}\right)p_\mu p_\nu=\Omega\tilde{\mathcal{H}},
\end{align}
where $V^\mu=\delta^{\mu}_0$ 
and $\Omega=A^{-4}a^{-2}$ is a conformal factor. In this last expression, we have introduced an effective refractive index $\tilde{n}$ which is given by the product of the material refractive index and a function of the metric, i.e., $\tilde{n}=(a A^3/B) n$. Note that for the class of metrics and the observer field that we are considering, the vacuum spacetime is equivalent to a medium with refractive index $n_{\rm sp}=a A^3/B$~\cite{de1971gravitational}. Thus, the effective refractive index $\tilde{n}$ is exactly the product of the spacetime refractive index $n_{\rm sp}$ and the material refractive index. {This is in line with the results in~\cite{spengler2023optical} obtained at the level of the full Maxwell's equations.}

We see that the Hamiltonian $\mathcal{H}$ is 
related\footnote{Via the factor $\Omega$ which is always positive as long as we are in the outer region of the central object.} to a Hamiltonian $\tilde{\mathcal{H}}$ representing the dispersion relation in flat spacetime for a medium with refractive index $\tilde{n}$ and characterized by the observer field $V^{\mu}$. The dispersion relation condition $\mathcal{H}=0=\tilde{\mathcal{H}}$ is the same {since the solutions of the Hamiltonian equations are left invariant by the multiplication of the Hamiltonian by a {nowhere vanishing}
factor\footnote{{For a transformation $\mathcal{H}(x,p)\rightarrow \tilde{\mathcal{H}}(x,p)= f(x,p)^{-1}\mathcal{H}(x,p)$ with everywhere positive function $f(x,p)$, the solutions $(x(s),p(s))$ to the Hamiltonian equations are invariant up to reparameterization {$ds\rightarrow d\tilde{s}=f(x(\tilde{s}),p(\tilde{s}))ds$ since $\tilde{\mathcal{H}}(x(\tilde{s}),p(\tilde{s}))=0$ (see also \cite{perlick2000ray}).}
}}.} Thus we expect the geometric optics to be the same in curved spacetime as in the effective medium in flat spacetime.

It is important to note that for the equivalence to hold the refractive index of the material needs 
to remain a function of the frequency $\omega$ and not a function of the frequency as defined in the effective medium in flat spacetime, i.e., $\nu=-{p}_\mu {V}^\mu$. 

\section{Blueshift in the presence of an RIP}
From now on, we focus on light propagating in a stationary medium in Schwarzschild spacetime. Thus we set the scale factor $a$ in Eq.~\eqref{metricspehrical} to one
and chose $A(r)=1+r_S/4r$ and $B(r)=1-r_S/4r$. This assumption simplifies the treatment of the Hamiltonian equations, and the corresponding redshift equation, while capturing the essential features of laboratory experiments in Earth's weak gravitational field. In particular, the redshift equation Eq.~\eqref{redshiftEq} simplifies to
\begin{equation}\label{eqred1}
    \frac{d (-p_a U^a)}{d{s}
    }=-(p_a U^a)^2\left[a_\mu{N}^\mu+\nabla_\mu n\,{U}^\mu\right] n, 
\end{equation}
where the first term encodes the gravitational redshift and the second term shows how the frequency is affected by a time-dependent refractive index. In the effective medium in flat spacetime description, the redshift equation assumes the form 
\begin{equation}\label{eqred2}
    \frac{d \nu}{d{\tilde{s}}
    }=-\nu^2\tilde{n}\partial_t\tilde{n},
\end{equation}
with {${ds}/{d\tilde{s}}=\Omega^{-1}$}.
{From the expressions in Eqs.~\eqref{eqred1} and~\eqref{eqred2}}
it is clear that, apart from the gravitational redshift, a time dependent refractive index of the material gives rise to variations in the frequency of the light. This effect is at the basis of the analogue models of white hole horizons investigated in the literature~\cite{philbin2008fiber,faccio2012optical,rubino2011experimental,PhysRevLett.105.203901,cacciatori2010spacetime,PhysRevD.83.024015}. 

Here we focus on how different gravitational gradients, in conjunction with the presence of an RIP originating from a propagating soliton, affect the frequency of weak probe pulses in non-linear media. In particular, we want to compare the blueshift experienced by a probe pulse, co-propagating with an RIP, moving outwards radially in Schwarzschild spacetime with the one of the same setup oriented horizontally. Note that, given the short propagation lengths that we will consider for the probe pulses (from a few millimeters to a meter), the horizontal propagation can be safely assumed as at constant radius. This configuration also lends itself to possible interferometric experiments sensitive to the difference in blueshift of the horizontal and radially propagating pulses.

\begin{figure*}
\centering
\includegraphics[scale=0.5]{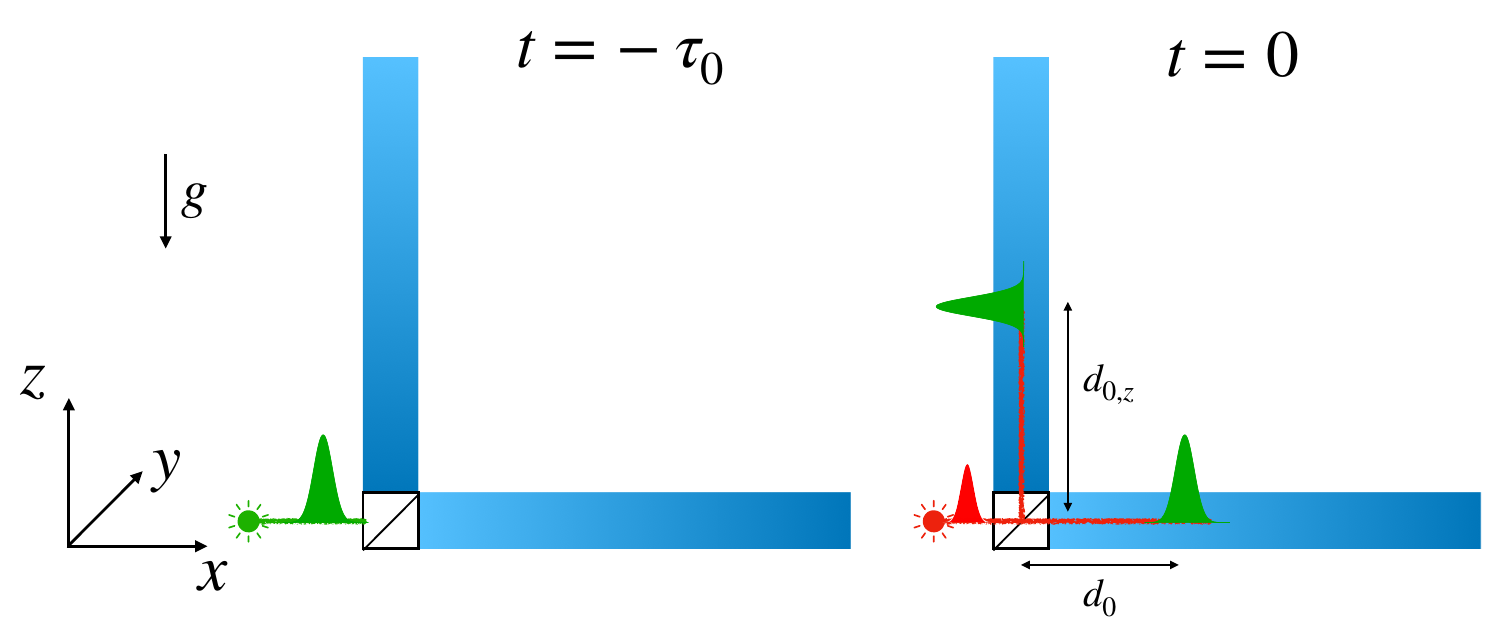}
    \caption{Schematic representation of the operational setup described in the main text. At coordinate time $t=-\tau_0$, a stationary observer at the origin of the coordinate system at $r=r_T$ {sends a strong light pulse that is split into two pulses. Each pulse enters a non-linear, dispersive optical fiber, one oriented horizontally (along $x$) and the other radially (along $z$). This leads to two propagating Gaussian RIPs in the fibers.}
 {The fiber oriented radially is attached at its top (not shown in the figure).} 
 At time $t=0$, a probe pulse is sent trailing the RIPs. {Finally, the frequency of the probe pulse is locally measured once the probe pulse is trailing the RIP well outside of the latter (several Gaussian widths ($>10\sigma$) distance from the center of the RIP).} 
    }
    \label{fig:scheme}
\end{figure*}

\subsection{Modelling the RIP}
A strong laser pulse in a non-linear dielectric modifies, through the nonlinear Kerr effect, the linear refractive index experienced by a weaker probe pulse and gives rise to a time-dependent RIP with profile
proportional to the pulse intensity~\cite{philbin2008fiber,rubino2011experimental}. In line with several works in the literature, we model the RIP profile $\delta n(t)$ as a Gaussian with constant width and peak-intensity parametrized by a parameter $\eta$ as
\begin{equation}\label{eq:deltan}
    \delta n(t)= \eta \exp\left(-\frac{|\vec{r}-\vec{r}_{\rm RIP}(t)|^2}{\sigma^2}\right),
\end{equation}
where $|\vec{r}-\vec{r}_{\rm RIP}|^2$ is the Euclidean distance between the spatial coordinates of the light ray ($\vec{r}$) -- the probe pulse -- and the center of the RIP following the trajectory $\vec{r}_{\rm RIP}(t)$ in the effective medium description in flat spacetime.
{The effective medium description allows us to use flat spacetime geometry when describing the trajectory of light rays which makes the description of the problem simpler. Another simplification comes from considering only the light ray trajectory collinear with the RIP, essentially reducing the problem to a one-dimensional one.}

The RIP is often assumed in the literature to propagate with a constant group velocity along the propagation direction, in which case the RIP trajectory reads $\vec{r}_{\rm RIP}=\{v_{\rm coord}\,t,0,0\}$. However, from the results in~\cite{spengler2023optical}, we know that for the radial propagation direction the velocity is not constant due to the gravity gradient (and mechanical stresses induced in the medium by gravitational forces). 

In~\cite{spengler2023optical}, we show that for the case in which the RIP propagates at (approximately) constant radius $r=r_T$, with $r_T$ being the radius of Earth, {the RIP velocity in the effective medium in flat spacetime is given by}
\begin{equation}\label{eq:velhor}
    v_{\rm coord}=\frac{1}{n_{\rm sp}}v,
\end{equation}
where $v$ is the constant propagation speed in the medium's {proper detector frame}\footnote{In general relativity, the proper detector frame is the frame determined by an orthonormal tetrad Fermi-Walker transported along the timelike trajectory of the physical system of interest which, in our setup, corresponds to a stationary observer.}. 
For the radial motion, in~\cite{spengler2023optical} we show that the velocity of the RIP in the effective spacetime medium is a linear function of the coordinate distance from $r=r_T$. {This linear dependence holds for realistic values of the parameters considered and can be safely assumed for short propagation lengths in Earth's gravitational field.} Thus, we parametrize the RIP velocity as {
\begin{equation}\label{eq:velvert}
    \tilde{v}=v_0+v_1 z,
\end{equation} 
where $v_0=v/{n_{\rm sp}(r=r_T)}$ coincides with the horizontal propagation case and $v_1=d\tilde{v}/dz|_{z=0}$}. 

By solving the equation $dt=dz/\tilde{v}(z)$ and then inverting the solution we find the RIP trajectory
\begin{equation}
    z(t)=\frac{v_0 e^{t v_1}-v_0}{v_1}\approx \frac{1}{2} t^2 v_0 v_1+t v_0,
\end{equation}
where, in the second equality, we have approximated the trajectory at the first order in $v_1 t\ll 1$. 
{Finally, as} discussed in~\cite{spengler2023optical}, the main factor affecting the RIP propagation speed in the radial direction is the presence of mechanical stresses that can change the refractive index of the medium making it a gradient-index one. These effects are modelled, in a realistic range of parameters, via photoelasticity due to the gravitational stresses on the medium due to its own weight {considering an optical fiber of length $L$ hanging attached at its top. In this case, stresses lead to a smaller density on the top than on the bottom, which reduces the refractive index at the top in comparison to the bottom and leads to an upward acceleration of the RIP}
{(see Appendix~\ref{appendixA1} and~\cite{spengler2023optical} for further details).}

\subsection{Operational setup}
Now that we know how to model the propagation of the RIP, we want to compare the blueshift sustained by a probe pulse that encounters the RIP propagating horizontally against the blueshift in the case of radial propagation. 
{In order to properly compare the blueshifts in these two situations, we consider a setup of two nonlinear, dispersive optical fibers, where one is oriented horizontally and the other radially (see Fig.~\ref{fig:scheme}). Furthermore, we consider a stationary observer in Schwarschild spacetime -- i.e., comoving with the fibers -- and positioned at $r=r_T$, with $r_T$ Earth's radius. This observer sends a strong light pulse to a beam splitter such that equally strong light pulses enter the two fibers at the same time  $t=-\tau_0$ (with $t$ the coordinate time). This leads to an RIP in each fiber. Then, after a coordinate time $\tau_0$, the same observer sends a probe pulse through the beam splitter following the RIPs, as depicted in Fig.~\ref{fig:scheme}.}

This configuration is such that, if we were in flat spacetime, the final blueshifted frequencies would be the same. Thus, we are interested in comparing the blueshift of the two probe pulses, once safely outside the respective RIP, when the experiment is performed in a weak gravitational field. {The difference in the blueshifted frequencies in the two propagation directions will encode {the effect of gravity }
in a similar way as the difference in the redshifted frequencies in the absence of the RIPs would. However, we will show that the RIPs lead to much stronger difference than gravitational redshift.} 

{Before proceeding, let us stress that} the physical reason why the probe pulses will be expelled by the region in which the RIP is present and afterwards lag behind the RIP is due to dispersion~\cite{cacciatori2010spacetime}. In the absence of dispersion, the probe pulse arrives to a standstill (in the  frame comoving with the RIP) at the phase horizon generated by the RIP and the frequency is infinitely blueshifted. Dispersion regularizes this unphysical situation: 
Since the frequency is strongly blueshifted, dispersion further slows down the probe pulse that will start to lag behind the RIP and keeps a constant frequency thereafter.

{For comparing the blueshifted frequencies, we numerically solve the Hamiltonian equations for the light rays
\begin{align}
    & \frac{dx^{\mu}}{ds}=\frac{\partial\tilde{\mathcal{H}}}{\partial p_\mu}\\
    & \frac{dp_{\mu}}{ds}=-\frac{\partial\tilde{\mathcal{H}}}{\partial x^\mu},
\end{align}
and then compute $\omega=-p_\mu U^\mu$ after the probe pulse is safely outside (typically several Gaussian widths) of the RIP. 
Given the previous discussion, the refractive index experienced by the probe pulse in the effective medium, and in the presence of the RIP, is given by 
\begin{equation}\label{fulln}
    \tilde{n}=n_{\rm sp}\left(n_0(\omega)
    -\frac{1}{2}n_0(\omega)^{3}\Delta(\varepsilon_r^{-1})+\delta n\right).
\end{equation}
This is the refractive index entering the Hamiltonian $\tilde{\mathcal{H}}$ and, in turn, 
the Hamiltonian equations and the redshift equation. In this expression, the refractive index experienced by the probe pulse, in the absence of the RIP, is given by the Cauchy formula 
\begin{equation}
    n_0(\omega)=A_{\rm Cauchy}+\frac{B_{\rm Cauchy}}{4\pi^2 c^2}
    \,\,\omega^2
\end{equation} 
that well describes the refractive index of common optical materials, e.g., silica, in the visible domain. The term containing $\Delta(\varepsilon_r^{-1})$ in Eq.~\eqref{fulln} encodes the effect of the photoelasticity on the probe pulse. Here photoelasticity is treated as a small perturbation to the inverse relative electric permeability $\varepsilon_r$ {and we consider $\Delta(\varepsilon_r^{-1})$ at first order in $r_S$}. Finally, $\delta n$ encodes the effect of the RIP as modelled in Eq.~\eqref{eq:deltan}. We refer the reader to Appendix~\ref{appendixA2} for additional details. }

{In what follows, we show
that the blueshift difference between the two propagation directions is not solely due to gravitational redshift and, in fact, has the opposite sign. Moreover, this difference is, in magnitude, between one and ten orders of magnitude greater than the gravitational redshift for realistic parameters. }

\begin{table}[]
\resizebox{\columnwidth}{!}{
\begin{tabular}{|ll|}
\hline
\multicolumn{2}{|l|}{{\textbf{Relative change in blueshifted frequency}}} \\ \hline
No photoelasticity & {$\Xi\approx 9.35\times 10^{-18}$} \\ \hline
Photoelasticity for the probe pulse &  {$\Xi\approx 4.25\times 10^{-9}$}  \\ \hline
Photoelasticity for probe pulse and RIP &  {$\Xi\approx 3.97\times 10^{-9} $}        \\ \hline
Photoelasticity for RIP only &  $\Xi\approx -2.76\times 10^{-10} $        \\ \hline
Gravitational redshift &  {$|\Xi_{\rm redshift}|\approx 1.42\times 10^{-18} $}        \\ \hline
\end{tabular}
}
\caption{Results of the numerical analysis using the parameters reported in Table~\ref{tab:parameters} in Appendix~\ref{appendixA4} and discussed in the main text. 
}
\label{tab2}
\end{table}

\begin{figure*}
\centering
\includegraphics[scale=0.5]{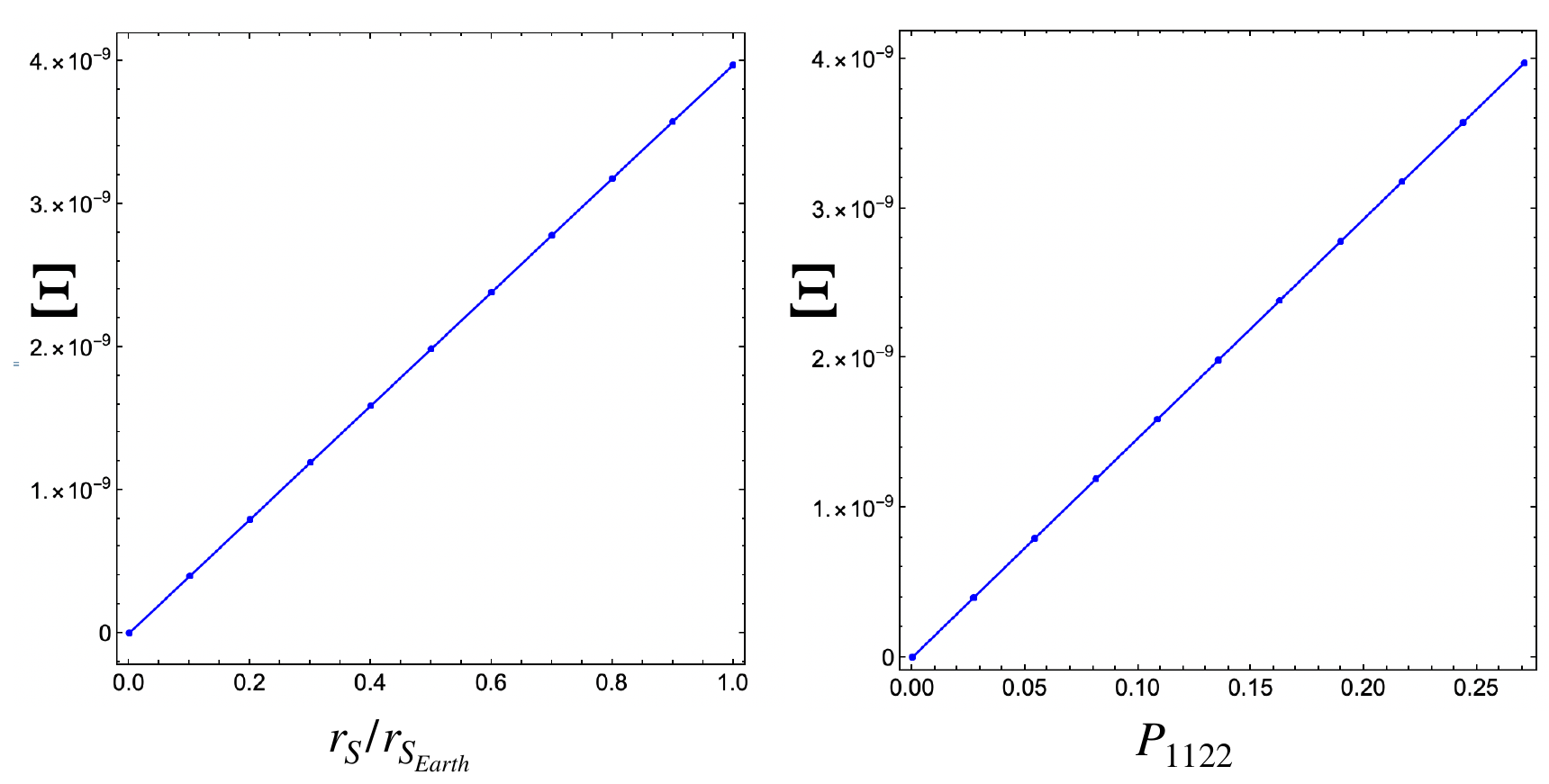}
    \caption{\textbf{Left panel:}
{Relative frequency shift 
 as a function of the Schwarzchild radius (in units of the Earth's Schwarzchild radius). Photoelasticity is accounted for in both the RIP and probe pulse.
 \textbf{Right panel:} Relative frequency shift 
 as a function of the photoelasticity parameter $P_{1122}$ for $r_S=r_{S_{Earth}}$. The value of all other parameters are reported in Table~\ref{tab:parameters} in Appendix~\ref{appendixA4} and discussed in the main text.
 }
    }
    \label{fig:loops}
\end{figure*}
\begin{figure*}
\centering
\includegraphics[scale=0.5]{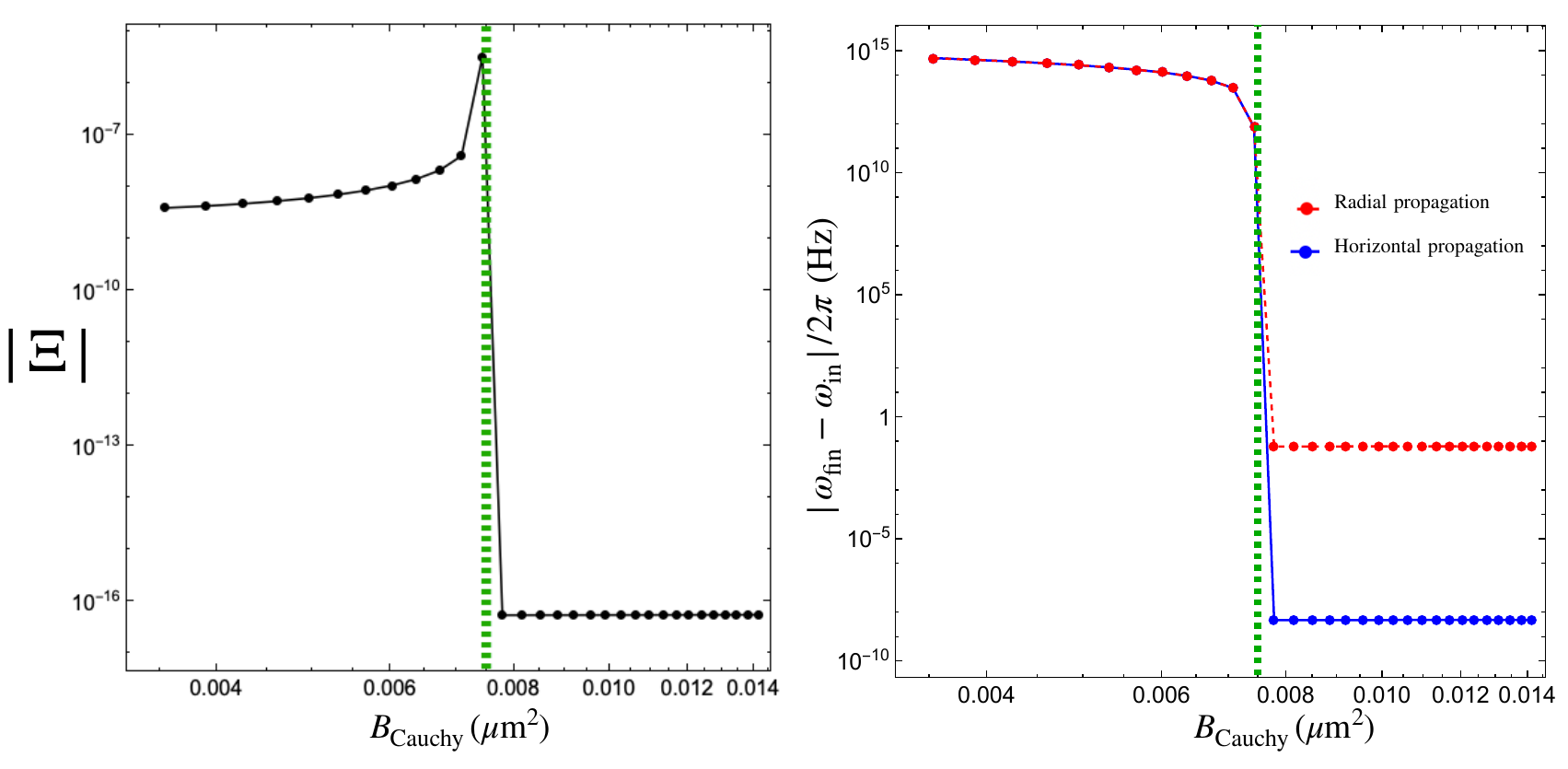}
    \caption{\textbf{Left panel:} Relative frequency shift between the two probe pulses (radial and horizontal propagation) as a function of the Cauchy constant $B_{\rm Cauchy}$ characterizing the material dispersion relation. \textbf{Right panel:} The absolute value of the differences between the initial and final frequency of the probe light as a function of the Cauchy constant $B_{\rm Cauchy}$ for both propagation directions.
    In both panels, the green dashed vertical line highlights that the amplification effect on $\Xi$ is strictly correlated with having a blueshift. Indeed, to the right of the green vertical line in the right panel $\omega_{\rm fin}-\omega_{\rm in}<0$, i.e., the probe light did not experience any blueshift. Photoelasticity is accounted for in both the RIP and probe pulse and all other material and geometrical parameters are given in Table~\ref{tab:parameters} in Appendix~\ref{appendixA4}.}
    \label{fig:dispersion}
\end{figure*}

\section{Results}
We have solved numerically the Hamiltonian equations for the Hamiltonian $\tilde{\mathcal{H}}$ for the light ray propagating in the same direction as the RIP and collinear with its center. While the problem is effectively one-dimensional, we nonetheless solve the equations fully in 3D restricting to the plane spanned by the $x$ and $z$ axes in Fig.~\ref{fig:scheme} 
{by setting $p_y(0)=0$ and $y(0)=0$} in the initial conditions (see Appendix~\ref{appendixA3}). 
{This allows us to justify the assumption that the horizontal propagation is, approximately, at constant radius. Indeed, in the absence of the RIP, we obtain a minuscule gravitational redshift of one part in $10^{-28}$ -- ten order of magnitude smaller than the redshift experienced by the radially propagated pulse and compatible with analytical results -- experienced by light due to the change in the radial distance in the horizontal propagation case.}

We define a relative frequency shift $\Xi$ as  
\begin{equation}
    \Xi=\frac{\omega_\perp-\omega_\parallel}{\omega_\perp+\omega_\parallel},
\end{equation}
where $\omega_\perp$ and $\omega_\parallel$ are the blueshifted frequencies in the radial and horizontal propagation directions, respectively. 
{We report the results of our numerical investigations in Table~\ref{tab2}. These results are obtained by choosing the parameters in Table~\ref{tab:parameters} in Appendix~\ref{appendixA4}. They correspond to the propagation of a soliton with central frequency $8.25\cdot 10^{14}$~Hz in fused silica, such that $v=0.65\,c$, Gaussian width $\sigma\approx 21$~$\mu$m, and $\eta=10^{-2}$ followed, after a coordinate time of approximately $8.65\cdot 10^{-13}$~s corresponding to a propagated coordinate distance for the horizontal RIP of $168$~$\mu$m, by a probe light pulse with frequency $\sim 5.69\cdot 10^{14}$~Hz in Earth's gravitational field. These values are in line with the ones reported in the literature on analogue Hawking radiation~\cite{cacciatori2010spacetime,PhysRevD.83.024015,faccio2010analogue,rubino2011experimental,PhysRevLett.105.203901}.}

Note that, when neglecting photoelasticity, (see first line of the table), 
we obtain a relative frequency shift that is approximately  
one order of magnitude greater than the (absolute value of the) gravitational redshift. 
If photoelasticity is included, parameterized by the value of the component for transverse stress of the photoelastic tensor of fused silica $P_{1122}=0.271$~\cite{biegelsen1974photoelastic,primak1959photoelastic},
the shift is ten orders of magnitude larger than the gravitational redshift. 
The second, third, and fourth row in Table~\ref{tab2} show 
that the main effect is given by the photoelasticity on the probe pulse and not by the photoelasticity-induced acceleration of the RIP radially propagating\footnote{{In~\cite{spengler2023optical}, it was also observed that, in radial propagation, the full width at half maximum of the soliton in the effective spacetime experiences a narrowing effect. Unfortunately, no analytical expression was obtained for such an effect {contrary to the case of the RIP velocity. Nevertheless, we have performed an order of magnitude estimate by using the same numbers as in~\cite{spengler2023optical} --- obtained for similar values of the physical parameters --- to include the narrowing  of the soliton's Gaussian width with the propagation distance. These consistency checks show that the  quantitative results in Table~\ref{tab2} are not significantly altered.}
}}.

\begin{figure*}
\centering
\includegraphics[scale=0.5]{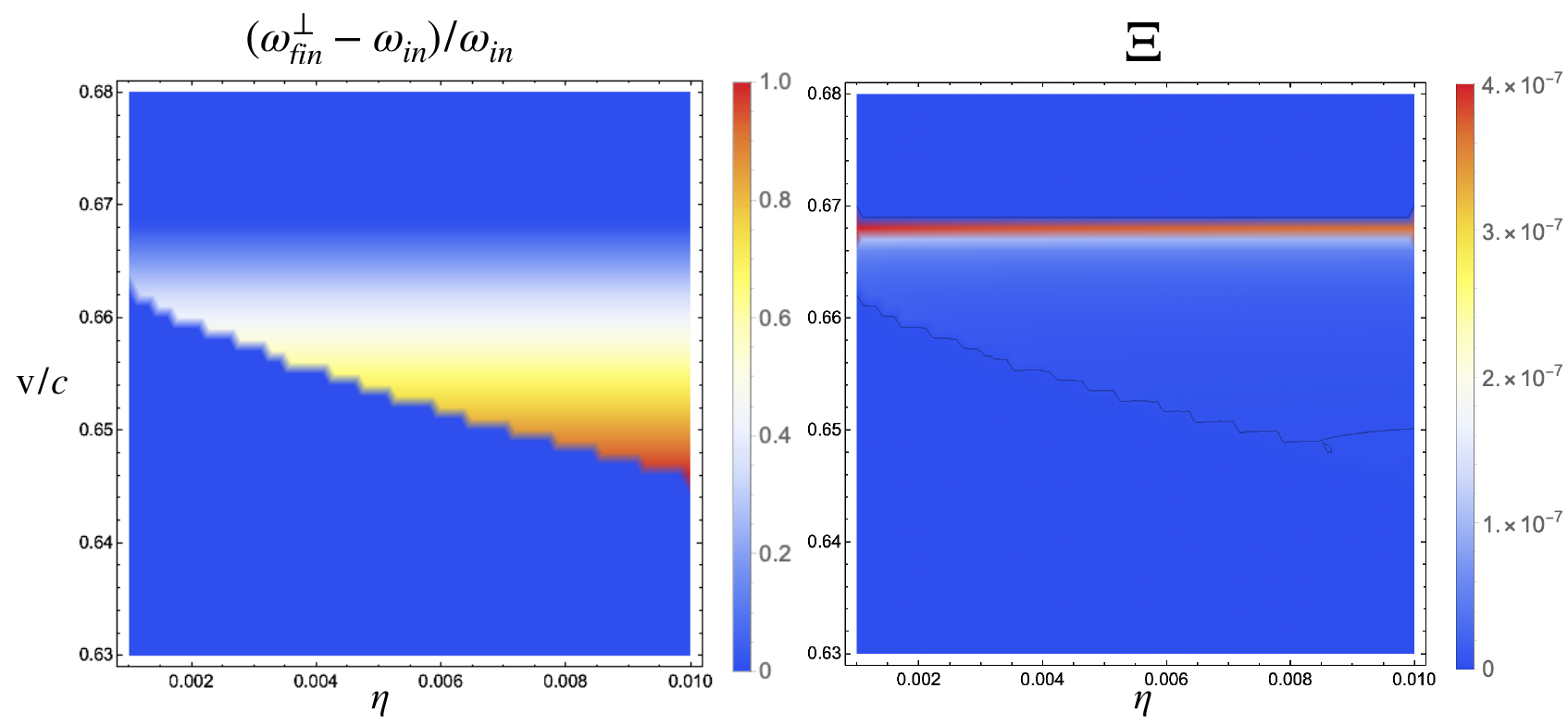}
    \caption{ 
    \textbf{Left panel}: Difference in the final and initial frequency for the radially propagating probe light as a function of the strength of the RIP $\eta$ and its velocity $v$. \textbf{Right panel:} Relative frequency difference as a function of the strength of the RIP $\eta$ and its velocity $v$ (note that the initial frequency of the probe light is $\approx 3.57\times 10^{15}\,{\rm (rad\cdot Hz)}$). The black contour represents the value for $\Xi$ reported in the third row of Table~\ref{tab2}. 
    In both panels, the non-smoothness of the contours is due to the discreteness of the grids of points that we have numerically analyzed (a total of 2321 points). Photoelasticity is accounted for in both the RIP and probe pulse and all other material and geometrical parameters are as given in Table~\ref{tab:parameters} in Appendix~\ref{appendixA4}.
    }
    \label{fig:density}
\end{figure*}
{In Fig.~\ref{fig:loops} we show, for the same material parameters, the linear dependence of the relative frequency difference $\Xi$ on the Schwarzchild radius and when varying the strength of photoelasticity.}

{Before proceeding to discuss the physical intuition behind the results just described, 
and in order to show that the results just discussed are not an artefact of the chosen parameters, we further explore the space of material and RIP parameters. Figure~\ref{fig:dispersion} shows the behavior of the relative frequency difference $\Xi$, as well as the overall blueshift in the horizontal and radial propagation cases, as a function of the parameter $B_{\rm Cauchy}$ entering the Cauchy formula for the dispersion relation of the medium. Note that, in the previous results $B_{\rm Cauchy}\approx 354\cdot 10^{-5}$~$\mu\rm{m}^2$ corresponds to the one of fused silica and we span one order of magnitude for  $B_{\rm Cauchy}$ which accounts for various common optical materials. From Figure~\ref{fig:dispersion} we see that $\Xi$ can be even larger than the one considered for fused silica for more dispersive material. Note that, for a fixed initial frequency of the probe light, increasing the dispersion can lead to the probe pulse never reaching the RIP. This is indeed what we observe in Figure~\ref{fig:dispersion} for $B_{\rm Cauchy}\geq 0.0075$~$\mu{\rm m}^2$.}
{Finally, Figure~\ref{fig:density} shows the inspection of the parameter space characterized by the RIP strength $\eta$ and the RIP velocity $v$ in the material proper detector frame. Note that modifying the RIP velocity amounts to changing its central frequency (see Appendix~\ref{appendixA1} for details). From Figure~\ref{fig:density} we can see once again that the amplification effect on $\Xi$ {does not hinge on a specific choice of parameters but is }
a fairly general feature that can be even larger than the value considered in Table~\ref{tab2} as long as photoelasticity is taken into account. This amplification effect is {large }
as far as the probe light actually experiences a blueshift due to the interaction with the RIP. Indeed, by comparing the right and left panels, we see that the amplification effect is absent or greatly diminished whenever no blueshift is registered\footnote{{Note that for small RIP velocities the probe pulses will traverse and overtake the respective RIPs. Nonetheless, in the radially propagating case, the probe pulse experiences a tiny blueshift effect due to the gradient-index nature of the effective medium. This effect is still some orders of magnitude greater than the expected gravitational redshift for the propagation distances considered. However, we see from Fig.~\ref{fig:density} that the corresponding $\Xi$ is negligible compared to the case in which a {large }
blueshift happens.}}.}

\subsection{Physical intuition}
We have seen that the difference of blueshifted frequencies in our setup is, in absolute value, between one and ten orders of magnitude greater than the gravitational redshift. While we obtained these results by directly solving the light ray equations, 
one can estimate the order of magnitude of the effect from considerations of momentum conservation. 

In~\cite{rubino2012soliton}, the authors consider the scattering process between a probe pulse 
in the form of a 
monochromatic wave and a polarization wave sourced by an RIP in flat spacetime. In the scattering picture, the blueshift is due to momentum exchange between the strong RIP pulse and the weak probe pulse. This exchange physically happens due to interaction terms in Maxwell's equations which are effectively captured by a changing refractive index profile. The conservation of the longitudinal momentum is formally derived in~\cite{rubino2012soliton} and is given by
\begin{equation}\label{eq:momentum}
    \kappa(\omega_{RR})=\kappa(\omega_{\rm in})+\frac{\omega_{RR}-\omega_{\rm in}}{v}, 
\end{equation}
where $\omega_{RR}$ is the blueshifted frequency, $\omega_{\rm in}$ is the input frequency of the probe pulse, $\kappa=\omega n(\omega)/c$ is the momentum (the dispersion relation) with $n(\omega)$ the medium refractive index without the RIP, and $v$ is the RIP constant velocity. Intuitively, Eq.~\eqref{eq:momentum} balances the change in momentum of the probe pulse with the momentum transferred by the RIP (last term on the RHS). Thus, in the flat spacetime analysis, in order to determine the blueshifted frequency due to the scattering it is enough to solve the system given by 
\begin{align}
    &\kappa(\omega)=\kappa(\omega_{\rm in})+\frac{\omega-\omega_{\rm in}}{v}\\
    &\kappa(\omega)=\omega n(\omega)/c,\nonumber
\end{align}
which gives the allowed modes in which the probe pulse can scatter. This picture also makes clear the crucial role of material dispersion which, 
by
making the dispersion relation cubic in the frequency, allows for a mode with blueshifted frequency\footnote{Turning off dispersion in the material has the, nonphysical, effect of moving the blueshifted mode frequency to infinity.}.

\begin{figure*}
\centering
\includegraphics[width=\textwidth]{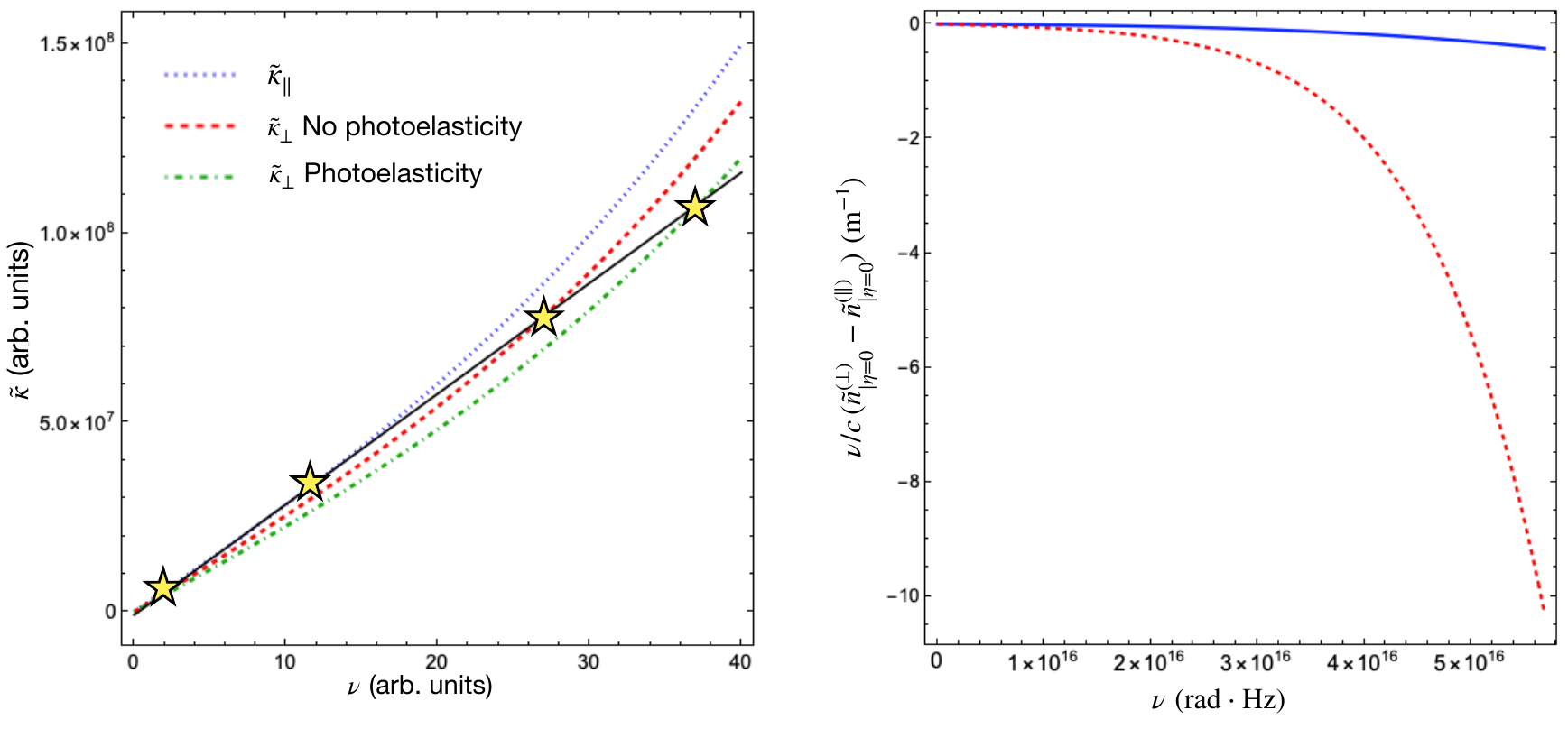}
    \caption{{\textbf{Left panel:} Sketch of the graphical solution of the system of equations describing momentum conservation and dispersion relation. {This sketch exaggerates, for the sake of visualization, the differences between the dispersion relations for the horizontal propagation ($\tilde{\kappa}_{\parallel}$, blue dotted curve), radial propagation with no-photoelasticity accounted for ($\tilde{\kappa}_{\perp}$, red dashed curve), and the radial propagation with photoelasticity ($\tilde{\kappa}_{\perp}$, green dot-dashed curve).}
    The solid, black line represents instead the momentum conservation relation. Finally, the stars represent the intersection between the curves and the solid line. The first intersection is the same for all. The blueshifted frequency is higher for the radial propagation and this effect is magnified further by photoelasticity. 
    \textbf{Right panel:} Difference between $\tilde{\kappa}_{\perp}$ and $\tilde{\kappa}_{\parallel}$ with no photoelasticity (solid, blue curve) and with photoelasticity (red, dashed curve), for the same parameters as reported in Table~\ref{tab:parameters} in Appendix~\ref{appendixA4}, as a function of the frequency $\nu$. Note that the photoleasticity magnifies the difference between the two dispersion relations as sketched in the left panel of this figure (which is not to scale).}
    }
    \label{fig:momentum}
\end{figure*}

In our case, the situation is slightly more complex than in flat spacetime. The flat spacetime analysis still applies unchanged, thanks to the effective medium formalism, for the horizontal propagation. {Indeed, in this case we can safely assume the propagation velocity of the RIP in the effective medium (Eq.~\eqref{eq:velhor}) as well as the refractive index $\tilde{n}_{|\eta=0}$ to be constant such that we have 
\begin{align}\label{eq:momconseff}
    &\tilde{\kappa}(\nu)=\tilde{\kappa}(\nu_{in})+\frac{\nu-\nu_{in}}{v_{coord}}\\
    &\tilde{\kappa}(\nu)=\nu \tilde{n}(\nu)_{|\eta=0}/c.\nonumber
\end{align}
Note that here we are using the frequency as defined in the effective medium, i.e., $\nu=-p_\mu V^\mu$ which is related to the physical frequency $\omega$ via the redshift factor of Schwarzschild spacetime.
} 
In the radial propagation case, however, we need to deal with a gradient-index medium in flat spacetime and an RIP with a non-constant velocity. {Indeed, the dependence of the spacetime refractive index $n_{\rm sp}$ and of the physical frequency $\omega$ on the radial coordinate makes the effective medium gradient-index. {Furthermore, the velocity of the RIP is not anymore constant but it is given by Eq.~\eqref{eq:velvert}.}
}

{The fact that in the radial propagation case we have a gradient-index medium and a non-constant propagation velocity of the RIP would call for the generalization of the results in~\cite{rubino2012soliton}.}
Nonetheless, {solving the system of equations in Eq.~\eqref{eq:momconseff} also for radial propagation}
still gives results in agreement with the ones obtained by numerically solving the light ray equations and sheds light on them (see also Appendix~\ref{App:new}). This is in accordance with the fact that the momentum conservation relations should be still approximately valid\footnote{{In the case in which the photoelasticity effect on the RIP is accounted for, to cope with a changing propagation speed of the RIP in solving the momentum conservation system of equations we resort to considering a constant averaged velocity defined as $\bar{\tilde{v}}=\frac{1}{r_{fin}}\int_0^{r_{fin}}dz (v_0+v_1 z)=v_0+v_1\frac{r_{fin}}{2}$.}}.
Moreover, it allows us to build some physical intuition on the working mechanism of the effect under consideration. By graphically solving the system of equations for the conservation of the longitudinal momentum and dispersion relation, for both propagation directions, we can see that (see Figure~\ref{fig:momentum}): 
i) when neglecting photoelasticity completely, the effect of curved spacetime is to make the effective medium gradient-index and such that {the graph of the dispersion relation as a function of the frequency $\nu>0$ for the radial propagation case is always below the one of the horizontal propagation case.}
Indeed, $n_{\rm sp}$ is a monotonically decreasing function of $z>0$ and, {furthermore}, the redshift factor in $\omega$ hinders the growth of the refractive index;
ii) when photoelasticity is accounted for, it has the same effect -- but magnified -- of increasing the frequency for the {intersection with the straight line corresponding to momentum conservation 
also thanks to a higher velocity of the radially propagating RIP in comparison to the horizontal one. 
{Looking back at Figure~\ref{fig:density}, we can now understand the fact that the magnitude of the blueshift
does not depend on the strength $\eta$ of the RIP. Indeed, this is expected from our intuitive picture given that $\eta$ does not appear in Eq.~\eqref{eq:momconseff}. The {serrated edge of the }
blueshift region in Figure~\ref{fig:density} is determined by the fact that, when $\eta$ is too small the probe pulse can penetrate the RIP and exit in front of it. In this case no blueshift is observed. The slower the RIP -- with respect to the fixed initial probe pulse propagation velocity -- the larger $\eta$ needs to be to observe the effect. The upper border of the blueshift region is instead determined by the fact that {above a certain RIP's propagation speed, the probe pulse is too slow to ever come to interact with the RIP.}}

\section{Discussion}
As we have discussed, experiments with weak probe pulses blueshifted by solitonic RIPs in non-linear media have been performed in the past, leading to the observation of interesting non-linear effects and the analogue of (stimulated) Hawking radiation. In this work, we have shown that when considering the presence of a weak gravitational field --- assuming a static, spherically symmetric spacetime metric ---, an interesting situation happens. Probe pulses propagating at constant radius and radially can be blueshifted such that the difference of their frequencies {is many orders of }
magnitude greater than gravitational redshift. 
{This is in particular the case when the mechanical stresses induced by gravity on the non-linear medium are accounted for.} 

We have performed an extensive exploration of the parameters space to show that {the amplification of the gravitational frequency shift  
} 
is quite general. Moreover, all the parameters that we have used in our numerical investigations account for experiments that can be performed in the lab with current technology. The effect appears for propagation distances of the order of few millimeters to a meter in common optical media. 

This opens the possibility of performing interferometric-like experiments to observe such gravitational effects. {We do not delve into the intricacies of how such experiments could be actually performed here since this is beyond the scope of this work. However, these experiments should aim at using coherent states of light passing through a beamsplitter as probe light and observing the interference between the probe light pulses after they are brought together via propagation in-vacuum --- completing the sketch in Fig.~\ref{fig:scheme} to a fully-fledged Mach-Zehnder-like interferometer.} 

While challenging, we believe such experiments could herald further investigations to detect {effects beyond Newtonian gravity}, {like gravitomagnetic effects,} in similar setups. {In this direction, more work to model the propagation of solitons in generic curved spacetime is needed.}

\section*{Acknowledgements}

We would like to thank Volker Perlick, Maria Chiara Braidotti and Daniele Faccio for useful discussions. 
AB and DB acknowledge support from the Horizon Europe EIC Pathfinder project QuCoM (Grant Agreement No. 10104697) and the Deutsche Forschungsgemeinschaft (DFG, German Research Foundation) project number BR 5221/4-1.  
D.R. acknowledges funding by the Federal Ministry of Education and Research of Germany in the project ``Open6GHub'' (grant number: 16KISK016) and support through the Deutsche Forschungsgemeinschaft (DFG, German Research Foundation) under Germany’s Excellence Strategy – EXC-2123 QuantumFrontiers – 390837967, and the TerraQ initiative from the Deutsche Forschungsgemeinschaft (DFG, German Research Foundation) – Project-ID 434617780 – SFB 1464.

\bibliography{references2.bib}

\clearpage
\onecolumngrid
\appendix

\section{Observation on material dispersion and analogue models}\label{appendix0}
{ 
The analogue gravity model given by light propagating in a non-linear dielectric medium in the presence of a refractive index perturbation induced by an intense laser pulse has been extensively investigated in the literature~\cite{rubino2011experimental,PhysRevLett.111.043902,rubino2012soliton,faccio2012optical,PhysRevLett.107.149402,PhysRevLett.105.203901,cacciatori2010spacetime,PhysRevD.83.024015,PhysRevLett.104.140403,faccio2010analogue}. In the dispersionless case, a Lorentzian metric can be identified with 
\begin{equation}\label{opticametricfaccio}
    \bar{g}_{ab}=g_{ab}+\left(1-\frac{1}{n(x)^2}\right)U_a U_b,
\end{equation}
where $U^a$ is the normalized four-velocity vector of the medium and $g_{ab}$ is the physical background metric (in the works cited this is assumed to be flat spacetime). As discussed in the main text, the frequency as measured by an observer in the rest frame of the medium is given by 
\begin{equation}
    \omega=-p_\mu U^\mu. 
\end{equation}
However, using the Hamiltonian equations we see that
\begin{equation}
    k^a\equiv\dot{x}^a=\frac{\partial \mathcal{H}}{\partial p_a}=\bar{g}^{ab}p_b,
\end{equation}
thus we can write $p_b=\bar{g}_{ab}k^a$ so that 
\begin{equation}
    \omega=-(p_a U^a)=-\bar{g}_{ab}k^b U^a=-\bar{g}(k,U).
\end{equation}
Then, the frequency measured by an observer in the rest frame of the medium is expressed as
\begin{equation}
    \omega=-\bar{g}(k,U), 
\end{equation}
with $k^a\equiv\dot{x}^a$ the tangent vector to the null geodesics of the optical metric. Note indeed that, in the dispersionless case, the dispersion relation $\mathcal{H}=0$ entails that $p_a$ are null vectors with respect to the optical metric and the Hamiltonian equations correspond to the geodesic Hamiltonian flow, meaning that the projections on the manifold coincide with null geodesics (given the condition $\mathcal{H}=0$) of the optical metric. 

In~\cite{cacciatori2010spacetime}, the authors include also the effect of dispersion in their treatment. Dispersion is indeed crucial to obtain physical results, i.e., to avoid non-physical infinite blueshifts. However, in~\cite{cacciatori2010spacetime}
dispersion is accounted for by adding the frequency dependence in the refractive index and proceeding in considering once again the geodesic equations without any change to the definition of the frequency expressed as $-\bar{g}(k,U)$. 

We notice that when dispersion is present the Hamiltonian equations do not correspond anymore to a geodesic Hamiltonian flow and no Lorentzian optical metric can be identified. Correspondingly, the definition of frequency as $-\bar{g}(k,U)$ does not coincide anymore with the physical one given by $-p_a U^a$. We should stress that this does not change the results in~\cite{cacciatori2010spacetime} at the qualitative level.
}

\section{RIP properties}\label{appendixA1}
{The Gaussian RIP we consider in the main text, while clearly an abstraction, can be thought of physically as a propagating soliton generated by a strong laser pulse in a Kerr non-linear material. While we consider the propagation of light in fused silica in the results in Table~\ref{tab2}, the following derivation can be also applied to other optical materials.}

{Let us start by considering the medium stationary in Schwarzschild spacetime in its proper detector frame. We model the dispersion relation of the medium via the Cauchy equation for the refractive index, keeping only terms up to order $\lambda^{-2}$, i.e., $n=A_{\rm Cauchy}+B_{\rm Cauchy}/\lambda^2$. Note that this parametrization of the dispersion relation has been used in~\cite{PhysRevD.83.024015} when dealing with a problem similar to the one we are considering here. As discussed in the same reference, the Cauchy equation is usually a good approximation in the visible spectrum and a more refined analysis could be made by using the full Sellmeyer formula for the dispersion relation. Nonetheless, at the current level of investigation, we deem the Cauchy equation a good compromise given its simplicity. 
}

{
For the pump pulse giving rise to the Gaussian RIP we assume a central frequency $\nu_{pulse}=2\pi c/\lambda_{pump}$ and expand the dispersion relation $\kappa=\nu n(\nu)/c$ around it 
\begin{equation}
    \kappa(\nu) =\kappa_0+\kappa _1 \left(\nu -\nu_{pulse}\right)+\frac{1}{2} \kappa _2 \left(\nu -\nu_{pulse}\right)^2.
\end{equation}
The first order term in this expansion encodes the group velocity of the RIP, i.e. its propagation velocity, in the medium's proper detector frame as $v=(\kappa_1)^{-1}$. This means that, fixing a certain value for $v$, and fixing a certain material, is tantamount to fixing the frequency of the pump pulse. 
}

As discussed in the main text, the propagation velocity of the RIP in the presence of a gravity gradient has been analyzed in~\cite{spengler2023optical}. There it is shown that the speed of the soliton in the effective medium picture, when moving at constant radial distance $r=r_T$, remains constant and is given by
\begin{equation}
    \tilde{v}=v/n_{\rm sp}(r_T).
\end{equation}
Furthermore, for radially outward propagating solitons, a numerical analysis of the non-linear Sch\"{o}dinger equation shows that the speed of the soliton in the effective medium is given by
\begin{equation}
    \tilde{v}=\left(\frac{\tilde{\kappa}_1}{1+z\frac{d\kappa_0}{dz}+\tilde{\kappa}_0^{-1}}\right)^{-1},
\end{equation}
where $\tilde{\kappa}_0=n_{\rm sp}\sqrt{-g_{00}}\kappa_0$ and $\tilde{\kappa}_1=n_{\rm sp}\kappa_1$ (see~\cite{spengler2023optical} supplemental material for further details).
{
Given that we are interested in the effect of weak gravitational fields and realistic parameters, the soliton velocity in the radial propagation case is approximately linear in the radial coordinate $z$ (see e.g. Fig.1 in~\cite{spengler2023optical}). Thus, we linearize the soliton velocity as 
\begin{equation}
    \tilde{v}=v_0+v_1 z,
\end{equation}
where $v_0=\frac{1}{n_{\rm sp}(r_T)}v$ is the constant velocity at constant radial distance $r=r_T$ characterizing the motion of the RIP at constant radius, and $v_1$ is given by 
\begin{align}
    v_1=& \frac{128 r_S r_T^2 \nu_{pulse} \left(\frac{c \kappa_1 (\kappa_0-\kappa_1 \nu_{pulse})}{\nu_{pulse}}+c \kappa_0 \kappa_2\right)}{c^2 \kappa_0 \kappa_1^2 (r_S+4 r_T)^4}
\end{align}    
}

{
However, as discussed in the main text and studied in detail in~\cite{spengler2023optical}, the main contribution to the speed of the vertically propagating soliton in our setup comes from including the effect of the mechanical deformations of the medium due to its weight in the gravitational field. These can be accounted for, in the range of realistic parameters that we are considering, by the photoelastic effect that modifies the refractive index of the medium making it a gradient-index medium. In particular, photoelasticity entails a perturbation of the electric permeability $\varepsilon_r\rightarrow \varepsilon_r+\Delta\varepsilon$. Considering a slab of material hanging in a weak gravitational field, i.e., a stationary slab of material, and limiting ourselves to isotropic materials and a diagonal stress tensor, the photoelastic perturbation can be written as
\begin{align}
    \Delta(\bm{\varepsilon}_r^{-1}) & = \frac{\mathcal{P}_{11\,22}}{Y}\sigma_{zz},
\end{align}
with $Y$ the Young module of the material and the stress on a material slab of cross area $A_\oslash$ given explicitly by
\begin{align}
    \sigma_{zz}(z)/Y &= \frac{F(z)}{A_\oslash} =  \frac{c^2}{c_s^2} \frac{r_S}{2r_0^2}\left( \frac{z}{\left(1-\frac{r_S}{4r_0}\right)\left(1+\frac{r_S}{4r_0}\right)^3} - \frac{z^2-2Lz}{r_0\left(1+\frac{r_S}{4r_0}\right)^6}\right)\\
    \mathcal{S}_{zz}(z) &= \frac{c^2}{c_s^2} \frac{r_S}{2r_0^2}\left( \frac{z}{\left(1-\frac{r_S}{4r_0}\right)\left(1+\frac{r_S}{4r_0}\right)^3} - \frac{z^2-2Lz}{r_0\left(1+\frac{r_S}{4r_0}\right)^6}\right),
\end{align}
where we used that the speed of sound in the fiber is $c_s = \sqrt{Y/\rho_m}$, $L$ is the total length of the slab of material, and $r_0=r_T+L$. For further details on the derivation of these expressions, we refer the interested reader to the supplemental material of~\cite{spengler2023optical}.
}

Photoelasticity modifies the linear refractive index of the material as $n(\omega)=\sqrt{1+\chi_1(\omega)+\Delta{\varepsilon}_r(\omega)}$. Since this enters in the definition of the $\kappa_i$ parameters which, in turn, {determine the velocity, we see that} photoelasticity affects the velocity 
of the soliton. Actually, only $v_1$ is affected and we report here the lengthy expression for completeness,
{\small
\begin{align}
    \frac{v_1}{c}=& \frac{128 r_S r_T^2 \nu_{pulse}}{c^2 \kappa_0 \kappa_1^2 (r_S+4 r_T)^4} \\ \nonumber
    &\left[\frac{8 c^7 \kappa_0^5 \kappa_1 P_{1122} (L+r_T)^2 (r_S-4 r_T) (r_S+4 r_T) \left(32 L (L+r_T) (r_S-4 (L+r_T))-(4 (L+r_T)+r_S)^3\right)}{c_s^2 \nu_{pulse}^5 \epsilon _r (r_S-4 (L+r_T)) (4 (L+r_T)+r_S)^6}\right.\\ \nonumber
    &\left.+\frac{8 c^5 \kappa_0^3 P_{1122} (L+r_T)^2 (r_S-4 r_T) (r_S+4 r_T) (2 \kappa_0-3 \kappa_1 \nu_{pulse})}{c_s^2 \nu_{pulse}^4 (4 L-r_S+4 r_T) (4 L+r_S+4 r_T)^6}\left(192 L^3+16 L^2 (r_S+28 r_T)\right.\right.\\ \nonumber
    &\left.\left. +4 L \left(3 r_S^2+16 r_S r_T+80 r_T^2\right)+(r_S+4 r_T)^3\right)\right. \\ \nonumber
    &\left.+\frac{c \kappa_1 (\kappa_0-\kappa_1 \nu_{pulse})}{\nu_{pulse}}+c \kappa_0 \kappa_2\right]\,.
\end{align}
}

\section{Probe pulse properties}\label{appendixA2}
{
The refractive index experienced by the probe light, in the absence of the RIP, is modelled once more by the Cauchy equation 
\begin{equation}
    n=A_{\rm Cauchy}+\frac{B_{\rm Cauchy} \omega ^2}{4 \pi ^2 c^2}
\end{equation}
with $B_{\rm Cauchy}$ accounting for material dispersion.
Furthermore, we can include the effect of photoelasticity also on the probe pulse via the relation
\begin{align}
    n&=\sqrt{\varepsilon_r+\Delta\varepsilon_r}\sim \sqrt{\varepsilon_r}+\frac{\Delta\varepsilon_r}{2\sqrt{\varepsilon_r}}\sim\sqrt{\varepsilon_r}-(\varepsilon_r)^{3/2}\Delta(\varepsilon_r^{-1})/2\\
    &=A_{\rm Cauchy} +\frac{B_{\rm Cauchy} \omega ^2}{4 \pi ^2 c^2}\,\omega^2-\frac{1}{2}\left(A_{\rm Cauchy} +\frac{B_{\rm Cauchy} \omega ^2}{4 \pi ^2 c^2}\,\omega^2\right)^{3}\Delta(\varepsilon_r^{-1}),
\end{align}
where we have used the fact that, for realistic values of the parameters, the correction to the refractive index due to photoelasticity is small and can be treated as a perturbation. Moreover, in our numerical analysis we consider the photoelasticity effect to first order in $r_S/r_T$. 
}

\section{Operational setup}\label{appendixA3}
Putting together all the elements that we have discussed so far, we have that the refractive index experienced by the probe light in the effective medium and in the presence of the RIP can be written as 
\begin{equation}
    \tilde{n}=n_{\rm sp}\left(A_{\rm Cauchy} +\frac{B_{\rm Cauchy} \omega ^2}{4 \pi ^2 c^2}\,\omega^2-\frac{1}{2}\left(A_{\rm Cauchy} +\frac{B_{\rm Cauchy} \omega ^2}{4 \pi ^2 c^2}\,\omega^2\right)^{3}\Delta(\varepsilon_r^{-1})+\delta n\right).
\end{equation}
This is the refractive index entering the Hamiltonian $\tilde{\mathcal{H}}$, the Hamiltonian equations, and the redshift equation. 

As stated in the main text, in order to properly compare the blueshifts in the two propagation directions, we consider a stationary observer in Schwarschild spacetime -- i.e., comoving with the physical medium in which light propagates -- and positioned at $r=r_T$. This observer sends two RIPs in the two propagation directions at $t=-\tau_0$ (with $t$ the coordinate time). Then, after a time $\tau_0$, the same observer sends a probe pulse following the RIPs. This operational setup allows us to set meaningful initial conditions for solving the system of Hamiltonian equations. Indeed, after a coordinate time $\tau_0$, the (center of the) horizontally propagating Gaussian RIP -- which we approximate as with constant velocity -- will be at a coordinate distance $d_0={v}_{0}\tau_0$ from the origin of the coordinate system that we are using. Note that for what concerns the results in Table~\ref{tab2} in the main text, we choose this distance to be a multiple of the Gaussian width of the RIP (specifically $8\sigma$) for the numerical analysis. In this same coordinate time $\tau_0$, the radially propagating soliton will have reached a different coordinate distance given by $d_{0,z}=d_0+\frac{d_0^2 v_1}{2v_{0}}$.

By fixing $d_0$, we can then determine the coordinate time $\tau_0$ needed for the horizontally propagating soliton to reach that distance and set that as the time at which the probe pulse is emitted. For the radially propagating solition, the initial coordinate position will be determined by $d_{0,z}$ and we use the trajectory
\begin{equation}
    z_{RIP}(t)=v_0 (t+\tau_0)+\frac{1}{2} v_0 v_1 (t+\tau_0)^2.
\end{equation}

Having determined $d_0$ and $d_{0,z}$, we can fix also the rest of the initial conditions. The probe pulses are assumed to start at the origin of the coordinate system which is chosen at $r=r_T$ in Schwarzschild coordinates. At $t=0$, the RIPs along the $x$ and $z$ axes have propagated a distance $d_0$ and $d_{0,z}$, respectively. The initial frequency is denoted as $\omega_0$ and set to $\omega_0=2\pi$ in units of $c/\lambda_0$ with $\lambda_0=527$~nm the probe light initial wavelength. For the propagation along the $x$-axis (the horizontal direction), we set $p_z(0)=p_y(0)=0$, $p_t(0)$ is determined by the initial frequency via $\omega_0=-p_t(0)\cdot U$. Finally, the dispersion relation $\tilde{\mathcal{H}}=0$ enforces the initial condition on $p_x(0)$. For the propagation along the $z$-axes (the radial direction), the initial conditions are obtained in the same way with, this time, $p_x(0)=p_y(0)=0$ and $p_z(0)$ enforced by the dispersion relation. 

\section{Physical intuition: comparison with the numerical results}\label{App:new}
{In the main text, we discussed how the blueshift phenomenology encountered can be better understood by using the results in~\cite{rubino2012soliton}. In the case of horizontal motion, the blueshifted frequency can be determined by graphically solving the system of equations
\begin{align}
    &\tilde{\kappa}(\nu)=\tilde{\kappa}(\nu_{in})+\frac{\nu-\nu_{in}}{v_{coord}}\\
    &\tilde{\kappa}(\nu)=\nu \tilde{n}(\nu)_{|\eta=0}/c.\nonumber
\end{align}
In the radial propagation case, the effective medium is a gradient-index one and the RIP propagation speed depends on the radial direction coordinate. This means that the equations above are, strictly speaking, not valid and a generalization should be sought. However, we have argued that these equations can still be used to find the blueshifted frequency phenomenology on the ground that we expect them to be approximately valid. In particular, to take care of the changing RIP velocity, we use a constant averaged velocity defined as $\bar{\tilde{v}}=\frac{1}{r_{fin}}\int_0^{r_{fin}}dz (v_0+v_1 z)=v_0+v_1\frac{r_{fin}}{2}$. Here we show in Figure~\ref{fig:comp} that indeed by doing so we recover the behaviour of the blueshifted frequency as shown in Figure~\ref{fig:density}. 
}

\begin{figure*}
\centering
\includegraphics[scale=0.5]{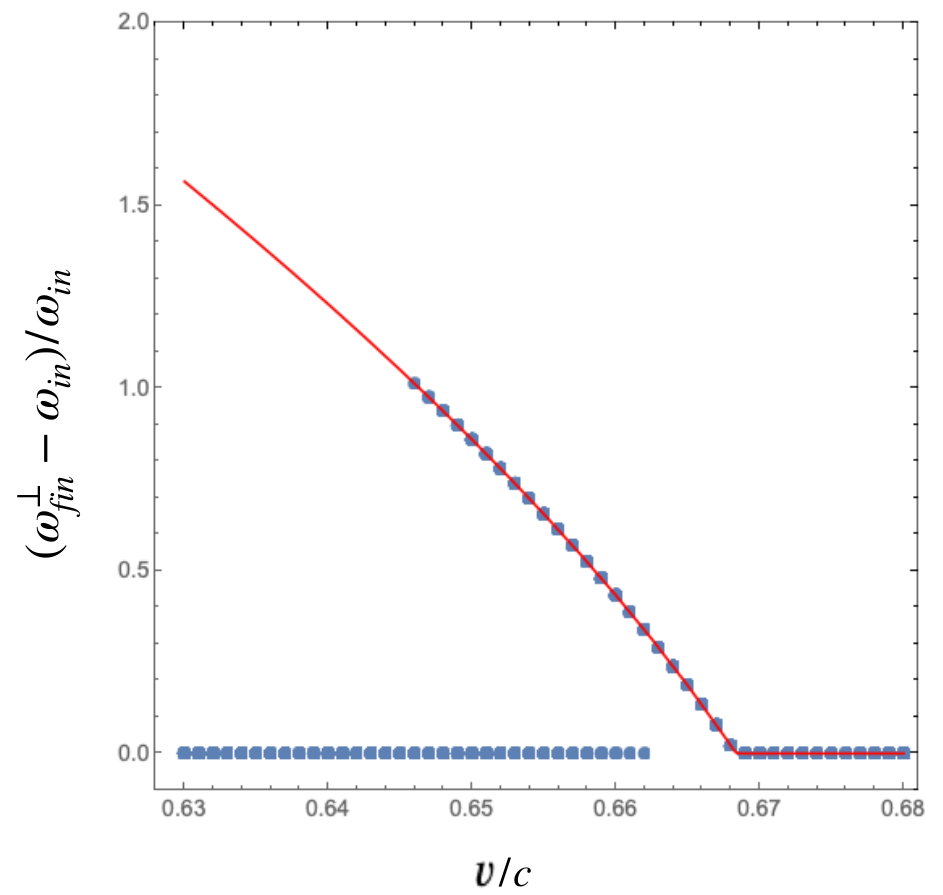}
    \caption{ 
    Comparison between the numerical simulations reported in Fig.~\ref{fig:density} left panel suppressing the $\eta$ dimension -- blue points -- and the prediction from solving graphically the system of equations Eq.~\eqref{eq:momconseff} using the average velocity discussed in the text -- red, solid curve. {The frequency difference is expressed as function of the initial speed of the RIP in the comoving frame of the medium ($v$) in units of $c$.}
    Photoelasticity is accounted for in both the RIP and probe pulse. We see that the red curve nicely interpolates the points whenever there is blueshift. The points that are zero on the left correspond to the cases in which the probe pulse penetrates and overtakes the RIP and the blueshifted mode is not excited.
    }
    \label{fig:comp}
\end{figure*}

\section{Numerical analysis parameters}\label{appendixA4}
Here we report the values of the parameters that we use to obtain the results in the main text as commented there.
\begin{table*}[h]
\centering
\resizebox{\textwidth}{!}{
\begin{tabular}{llll}
    \hline
    \textbf{Symbol}  & & \textbf{Name} & \textbf{Value/Expression}  \\ \hline\hline
    Properties of the medium (fused silica): \\ \hline
    ${P}_{11\,22}$ & & Component for transverse stress of the photoelastic tensor from \cite{biegelsen1974photoelastic,primak1959photoelastic} & 0.271 \\
    $c_s$ & & Speed of sound tabulated in \cite{heraeusfusedsilica} & 5720~m/s \\
    $L$ & & Total length of the medium  & between $0.1$ and $10$~m \\\hline
    Soliton pulse:\\ \hline
    $\sigma$ & & Gaussian pulse width  & $\sim 20\,\mu$m \\
    $\eta$ & & RIP parameter  & $10^{-2}$ \\
   $\lambda_{pulse}$ & & Central soliton wavelength & 363\,nm \\
   $v=(\kappa_1(\nu_{pulse}))^{-1}$ & & Soliton speed's initial condition & 0.65\,$c$ \\
    \hline
    Probe pulse:\\ \hline
  $\lambda_0$ & & Probe pulse's initial wavelength & $527$~nm \\
    $A_{\rm Cauchy}$ & & Cauchy equation's constant & $1.458$ \\
  $B_{\rm Cauchy}$ & & Cauchy equation's constant & $354\cdot 10^{-5}$~$\mu\rm{m}^2$ \\ \hline
    Miscellaneous:\\ \hline
    $r_\oplus$&& Earth equatorial radius&6378137\,m\\
    $r_S$ (Earth) && Schwarzschild radius of Earth & $9 \cdot 10^{-3}$\,m\\
\end{tabular}
}
\caption{Specifics of all the parameters entering the numerical simulations. }
\label{tab:parameters}
\end{table*}

\end{document}